 \documentclass[twocolumn]{aastex61}
\hypersetup{linkcolor=red,citecolor=blue,filecolor=cyan,urlcolor=magenta}

\usepackage{natbib}

\usepackage{gensymb}

\usepackage{amsmath} 
\newcommand{\angstrom}{\text{\normalfont\AA}}
\accepted{\today}
\shorttitle{Rotational Modulations in a T8 Dwarf Companion}
\shortauthors{Manjavacas et al.}
\begin{document}

\title{Cloud Atlas: Rotational Spectral Modulations and \textit{potential} Sulfide Clouds in the Planetary-mass, Late T-type Companion Ross 458C}

\correspondingauthor{Elena Manjavacas}
\email{emanjavacas@keck.hawaii.edu}

\author{Elena Manjavacas}
\affil{W. M. Keck Observatory, 65-1120 Mamalahoa Highway, Kamuela, HI 96743, USA}
\affil{Department of Astronomy/Steward Observatory, The University of Arizona, 933 N. Cherry Avenue, Tucson, AZ 85721, USA}

\author{D\'aniel Apai}
\affiliation{Department of Astronomy/Steward Observatory, The University of Arizona, 933 N. Cherry Avenue, Tucson, AZ 85721, USA}
\affiliation{Department of Planetary Science/Lunar and Planetary Laboratory, The University of Arizona, 1640 E. University Boulevard, Tucson, AZ 85718, USA}
\affiliation{Earths in Other Solar Systems Team, NASA Nexus for Exoplanet System Science}

\author{Ben W. P. Lew}
\affiliation{Department of Planetary Science/Lunar and Planetary Laboratory, The University of Arizona,
1640 E. University Boulevard, Tucson, AZ 85718, USA}

\author{Yifan Zhou}
\affiliation{Department of Astronomy/Steward Observatory, The University of Arizona, 933 N. Cherry Avenue, Tucson, AZ 85721, USA}

\author{Glenn Schneider}
\affiliation{Department of Astronomy/Steward Observatory, The University of Arizona, 933 N. Cherry Avenue, Tucson, AZ 85721, USA}

\author{Adam J. Burgasser}
\affiliation{Center for Astrophysics and Space Science, University of California San Diego, La Jolla, CA 92093, USA}

\author{Theodora Karalidi}
\affiliation{Department of Astronomy and Astrophysics, University of California, Santa Cruz, California, USA}

\author{Paulo A. Miles-P\'aez}
\affiliation{The University of Western Ontario, Department of Physics and Astronomy, 1151 Richmond Avenue, London, ON N6A 3K7, Canada}
\affiliation{Department of Astronomy/Steward Observatory, The University of Arizona, 933 N. Cherry Avenue, Tucson, AZ 85721, USA}

\author{Patrick J. Lowrance}
\affiliation{IPAC-Spitzer, MC 314-6, California Institute of Technology, Pasadena, CA 91125, USA}

\author{Nicolas Cowan}
\affiliation{Department of Earth \& Planetary Sciences, 3450 University St. Montreal, Quebec H3A 0E8, Canada}

\author{Luigi R. Bedin}
\affiliation{INAF — Osservatorio Astronomico di Padova, Vicolo Osservatorio 5, I-35122 Padova, Italy}

\author{Mark S. Marley}
\affiliation{NASA Ames Research Center, Mail Stop 245-3, Moffett Field, CA 94035, USA}

\author{Stan Metchev}
\affiliation{The University of Western Ontario, Department of Physics and Astronomy, 1151 Richmond Avenue, London, ON N6A 3K7, Canada}

\author{Jacqueline Radigan}
\affiliation{Utah Valley University, 800 West University Parkway, Orem, UT 84058, USA}





\begin{abstract}

Measurements of photometric variability at different wavelengths provide insights into the  vertical cloud structure of brown dwarfs and planetary-mass objects.  In seven {Hubble Space Telescope} consecutive orbits, {spanning $\sim$10~h of observing time}, we obtained time-resolved spectroscopy of the planetary-mass T8-dwarf Ross~458~C using the near-infrared Wide Field Camera 3.  We found spectrophotometric variability with a {peak-to-peak signal of 2.62$\pm$0.02~\%} ({in the 1.10-1.60~$\mu$m white light curve}). Using three different methods, we estimated a rotational period of 6.75$\pm$1.58~h for the white light curve, and similar periods for  narrow $J$- and $H$- band light curves. Sine wave fits to the narrow $J$- and $H$-band light curves suggest a {tentative} phase shift between the light curves with wavelength {when we allow different periods between both light curves. If confirmed, this phase shift may be similar to the phase shift detected earlier for the T6.5 spectral type 2MASS J22282889--310262}. We find that, {in contrast with   2M2228}, the variability of Ross~458C shows evidence for a {color trend} within the narrow $J$-band, but gray variations in the narrow $H$-band. The spectral time-resolved variability of Ross~458C might be {potentially} due to heterogeneous sulfide clouds in the atmosphere of the object. Our discovery extends the study of spectral modulations of condensate clouds to the coolest T dwarfs, planetary-mass companions.

\end{abstract}




\section{Introduction} \label{sec:intro}

At effective temperatures {close to} 1400~K -- within a relatively narrow range  --  brown dwarf upper atmospheres undergo a radical transformation that marks the {transition between the L and T spectral classes}. In the near-infrared spectra of T dwarfs, methane absorption features appear and near-infrared colors shift dramatically to bluer  \citep{Burgasser2002,Cushing,Kirkpatrick}. Below this temperature,  silicate and iron clouds seem to play a less important role {in shaping the emerging spectra} than for L-type dwarfs. This transition is thought to be due to clouds sinking below the photosphere \citep{Ackerman_Marley2001, Burgasser2002b}. 

Time-resolved photometry and spectroscopy allow isolating changes in cloud properties from bulk properties of the objects. Models of spectrophotometric variability in T-dwarfs like the T2 \object{SIMPJ013656.5+093347} (SIMP0136) \citep{Artigau2009,Apai2013},  the T2.5 \object{2MASS J21392676+0220226} (2M2139) \citep{Radigan2012,Apai2013},  and the T6.5 \object{2MASS J22282889--4310262} (2M2228) \citep{Buenzli2012} show that cloud layers {of} varying {thicknesses} are present in T-dwarfs atmospheres. Extensive monitoring of early T-type brown dwarfs demonstrated that the cloud thickness is constantly modulated by planetary-scale waves \citep{Apai2017}, and that continuously evolving light curves are very common  in brown dwarfs. The large-scale waves are possibly caused by feedbacks between the cloud layer and atmospheric dynamics: one-dimensional models {found} oscillations caused by latent heat \citep[][]{TanShowman2017}, and three-dimensional models found small- and large-scale waves and oscillations in coupled atmospheric dynamics and cloud evolution systems \citep{ShowmanTan2018}. The picture that emerged for these T-dwarfs is probably valid for most T-dwarfs: \citet{Metchev2015} found that $36^{+26}_{-17}\%$ of the T-dwarfs in their 16 object sample observed with Spitzer {at 3.6 and {4.5}~$\mu$m} are variable with  amplitudes between 0.8\% and 4.6\%. This finding indicates that clouds are typical to T-dwarf photospheres. 

Similarly, photometric variability probably due to heterogeneous cloud coverage has been found in unbound late-L and T planetary-mass objects and companions to stars. \cite{Biller2015} found {$J$-band} high-amplitude photometric variability in PSO~J318.5−22, an L7 dwarf with an estimated mass of $\sim$8~$\mathrm{M_{Jup}}$. In addition, \cite{Biller2018} reported phase shifts between the Spitzer {InfraRed Array Camera (IRAC)} light curve and those derived from the near-infrared {Hubble Space Telescope} (HST), {Wide Field Camera 3} (WFC3) spectra. Other examples of photometric or spectrophotometric variability found for planetary-mass objects include GU Psc b, a 9-13~$\mathrm{M_{Jup}}$ T3.5 companion with $\sim$4\% of variability amplitude \citep{Naud2017}; 2M1207b, a 2.3-4.8~$\mathrm{M_{Jup}}$ L5 companion, with variability amplitude up to 1.36\% \citep{Zhou2016}; 2MASS J11193254-1137466AB and WISEA J114724.10-204021.3, two L7 with masses of 4--6~$\mathrm{M_{Jup}}$, and variability amplitudes up to $\sim$2\% and $\sim$1\% in the [3.6] and [4.5] Spitzer channels, respectively \citep{Schneider2018}. Furthermore, two of the brown dwarfs (SIMP0136 and \object{2MASS J13243553+6358281}) for which \citet{Apai2017} reported planetary-scale {wave-modulated} cloud thickness {variations}, are also likely in the planetary-mass range (\citealt[]{Gagne2017,Gagne2018}).

Small amplitude rotational modulations appear also to be common in Y dwarfs. \cite{Cushing2016} has discovered photometric variability in  WISE J140518.39$+$553421.3, a Y0.5 brown dwarf with an estimated mass between 9--21~$\mathrm{M_{Jup}}$ \citep{Leggett2017}. \cite{Cushing2016} measured a variability amplitude up to 3.5\% in the  [3.6] and [4.5] Spitzer channels. They found that the amplitude was different in two epochs, leading them to conclude that the cloud structures might evolve with time. \citet[][]{Leggett2016} has also detected photometric variations in another Y0-dwarf (\object{WISEP J173835.52+273258.9}) in the Spitzer/IRAC bands and also identified likely near-infrared variability. {Finally, \cite{Esplin2016} found up to 5\% variability on the coldest know brown dwarf WISE J085510.83$-$071442.5 (Y2) at the [3.6] and [4.5] Spitzer bands. }

In conclusion, these examples show that clouds with varying vertical structures are also present in the coolest and lowest mass brown dwarfs and planetary-mass objects; but most of our knowledge currently is derived from L/T transition objects, that are relatively bright and most likely to have high-amplitude modulations \citep[]{Radigan2014}.

For Ross~458C, {\cite{Burgasser2010b}, \cite{Burningham2011}, and \cite{Morley2012} suggested that it should have a cloudy atmosphere as well. Nevertheless, \cite{Metchev2015} observed Ross~458C with the Spitzer $[3.6]$ and  $[4.5]$ channels during a 21~h-long continuous observation, concluding that its variability {was} not higher than 1\% at those wavelengths.}
In this work, we show the existence of spectrophotometric variability in the very late T-spectral type Ross~458C in time-resolved spectra acquired with the Wide Field Camera 3 (WFC3) onboard the Hubble Space Telescope (HST) during seven {consecutive spacecraft} orbits.

\section{Ross 458C}\label{Ross458C}

Ross~458C (R.A. 13:00:41.15, Decl. +12:21:14.22) is a T8 spectral type brown dwarf, with $J_{UKIDSS}$ = 16.69$\pm$0.01, and anomalously {red in the near-infrared}   ($J-K$ = $-$0.21$\pm$0.06). Ross~458C was discovered as a companion to a high proper motion binary system, Ross~458AB, in the Data Release 5+ of the UKIRT Deep Infrared Sky Survey (UKIDSS) Large Area Survey \citep{Goldman2010, Scholz2010}. 
Ross~458AB is a M0.5/M7.0 binary system at 11.51$\pm$0.02~pc \citep{Gaia2018}. Ross~458A is very active, and shows strong H$\alpha$ emission and photometric variability \citep{Hawley1997}, indicating a likely maximum age of the system of 400--800~Myr \citep{West2008}. 

\cite{Burgasser2010b} used a $V$ - $K/M_{K}$ color magnitude diagram  to estimate the metallicity of Ross~458A \citep{Johnson2009,Schlaufman2010}, and thus the metallicity of the system. They obtained a metallicity between [Fe/H] = $+$0.31$\pm$0.05, and [Fe/H] = $+$0.20$\pm$0.05, for  the metallicity calibrations of \cite{Johnson2009}, and \cite{Schlaufman2010} respectively. Both calibrations, thus, argue for supersolar metallicity.
\cite{Burgasser2010b} did not find Li absorption at 6708~$\angstrom$  in {the red optical spectrum of Ross~458A}, setting 30--50~Myr as a lower limit for the age of the system. Consistently with this estimate, the equivalent widths of the alkali lines in Ross~458AB are {larger} {than} those stars of similar spectral types in the Pleiades open cluster (112$\pm$5~Myr, \citealt{Dahm2015}), indicating that Ross~458AB is older than Pleiades. At the same time, Ross~458A and B are not tidally locked \citep{West2008}, implying an age lower than $\sim$1~Gyr. The BANYAN~$\Sigma$ tool \citep{Gagne2018}, {which} estimates the probability of membership to young moving groups using the kinematics of the targets, provides a 99.3\% of probability for the Ross~458ABC system belonging to the Carina Near Moving Group, with an estimated age of 200$\pm$50~Myr \citep{Zuckerman2006}.

\section{Observations and Data Reduction}

Ross~458C was observed in Cycle 23 of the HST program (PI D. Apai, GO-14241) using the Wide Field Camera 3 (WFC3) in its near-infrared channel and G141 grism \citep{MacKenty2010}. The WFC3/G141 grism covers the wavelength range between 1.05 and 1.70~$\mu$m, with a spectral resolving power of 130 at 1.4 $\mu$m. WFC3/IR has an {image} scale of 0.13 $^{\prime\prime}$/pixel. 

We acquired seven consecutive orbits of observations on 2018 January 6. {In each $\sim$95~min long orbit during the uninterrupted target visibility periods} we obtained eleven {G141} frames with an integration time of 201.4~s each. {To obtain an accurate wavelength reference for wavelength calibration}, four direct {(spectrally non-dispersed)} images in each orbit were also taken in the F132N filter. We used a 256$\times$256 subarray mode to eliminate {intra-orbit dead-time due to readout overheads}. We performed the data reduction using the same method as in previous works published by our group \citep[e.g.,][]{Apai2013,Buenzli2014,Buenzli2015,Lew2016,Manjavacas2018,Zhou2018}. 

{We restricted the spectral range studied to 1.20--1.32 and 1.54--1.60~$\mu$m to avoid the noise at the edges of the spectra (due to the drop in the instrument sensitivity) and in the 1.40~$\mu$m water band (where the source flux is very low). After reduction the average spectrophotometric uncertainty in measured intensity per spectral bin is 0.34\%}. The measured uncertainties are due to photon and read noise, plus potential residual systematic errors of recognized origins that we discuss in Section \ref{systematics}.

\section{Systematics Assessment}\label{systematics}

\subsection{Ramp effect correction}

The most prominent systematics in WFC3 near-infrared {time-resolved photometry} is the ramp effect. We corrected this systematic effect by using the process described in \cite{Zhou2017}, that models charge trapping and delayed release in the detector. {This} reproduced the ramp effect {with very high fidelity}, allowing us to correct for it.

\subsection{Pointing {stability}}

{High-precision differential spectrophotometry requires sub-pixel line-of-sight pointing stability. Such performance is usually delivered by the HST pointing control system but, infrequently, anomalously high "pointing drifts" can arise.  We evaluated, and confirmed, the requisite level of pointing stability during our Ross 458C observations as follows for each of the 77 spectra taken during the seven HST orbits. We measured the positions the photocentric peaks of the spectra in the cross-dispersion direction by Gaussian profile fitting at six equally separated  wavelength locations along the spectra.  We found that the dispersion in Gaussian-fit peaks in the image ensemble was at most 10\% of a pixel (13~mas)}. This is the nominal value for the observations and it is much smaller than the aperture size (4 pixels) that we used to extract the spectra. Thus, we conclude that the pointing instability does not affect significantly our measurements.

\subsection{Sky background variations}

Ross~458C is a companion to the Ross~458AB binary system with an AB--C separation of 102". This separation is much larger that the field of the WFC3 near infrared detector (33"$\times$33" in the 256$\times$256 subarray), therefore, we do not expect the flux of the binary to contaminate the flux of Ross~458C. Nevertheless, we tested if the flux of the sky (after sky subtraction) is correlated with the spectral-photometric variability found in the light curve of Ross~458C. We measured the flux of the sky as a function of time in {one} rectangular aperture of 146$\times$8  pixels above (x = 447--593, y = 552--560) and {one} below (x = 447--593 y = 513--521) the spectral trace. 

We calculated the $\tau$ Kendall's coefficient and its significance to quantitatively test for a correlation between sky background measurements in the two regions and the variability measurements of the white light curve {(1.10-1.60~$\mu$m)} {in the 77 images}, using the {\texttt{r\_correlate.pro IDL} function}. For the sky measured in the upper region of the Ross~458C spectra, we obtained a $\tau$ = $-0.08$ with a significance of 0.33, and for the region below the target's spectra we found a correlation of $\tau$ = $-0.02$, with a significance of 0.75. Both $\tau$ values are close to 0, indicating no correlation between  the target's spectro-photometric variability and the sky background.

\section{Results}

In Figure \ref{sinefit_LC} we present the white, the narrow $J$-band {(JNB, 1.21-1.32~$\mu$m)} and narrow $H$-band light curves {(HNB, 1.54-1.60~$\mu$m)} for Ross~458C, after charge trapping correction. All three light curves show quasi-periodic variations.


\subsection{{Spectral Variability}}

\begin{figure}
  \includegraphics[width=\linewidth]{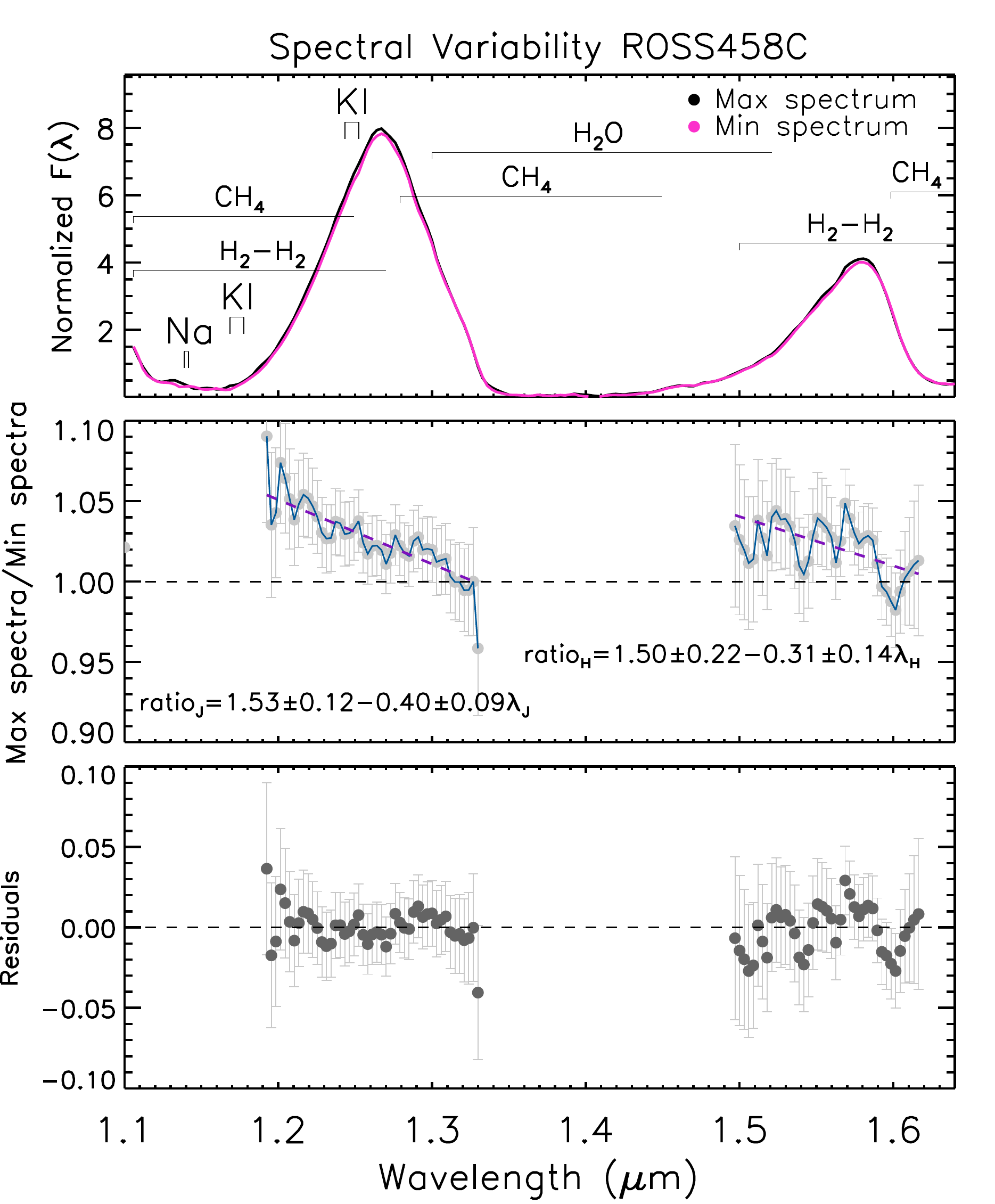}
  \caption{Top panel: average of the three maximum ({purple} color) and the three minimum ({black} color) spectra of the 77 spectra taken during seven HST orbits for Ross~458C. {Mid panel: ratio between the three spectra with the maximum flux and three spectra with the minimum flux of Ross~458C. We found a best linear fit to the J- and H-band ratios to show the color trend found in both bands. Bottom plot: Residuals of the ratio between the three maximum and minimum spectra after sustracting the best linear fit for both bands.}}
  \label{max_min}
\end{figure}

Following \cite{Apai2013}, we study the amplitude of the rotational modulations as a function of wavelength by comparing the average of the three maximum and  of the three minimum spectra among the 77 spectra  taken (see Figure \ref{max_min}, upper plot). In the lower panel of Figure \ref{max_min} we show the ratio between the three brightest spectra and the three faintest spectra -- in other words, the relative amplitude across the spectral wavelength {range}. 

The relative {peak-to-peak} signal of the rotational modulation for Ross~458C is 2.62$\pm$0.02~\% as measured {in the white and in the $JNB$ light curve. We measured a marginal rotational modulation of  3.16$\pm$1.36\% (2.2$\sigma$) in the $HNB$ light curve}. Because in several wavelength ranges of the spectra the flux density is close to zero (1.10--1.20 $\mu$m, 1.32--1.43 $\mu$m, and 1.63--1.69 $\mu$m), the ratio of these flux densities is very noisy. Therefore, we cannot derive any conclusion about the amplitude of the variations inside and outside the water band (1.35--1.43~$\mu$m). Outside the wavelength ranges (with signal-to-noise bigger than 20), a  {color trend (larger amplitude variations at shorter wavelengths) is visible in the narrow $J$-band, as shown by the best linear fit (Fig. \ref{max_min} mid panel), and also to some extent in the H-band, although the latter color trend is less significant (Fig. \ref{max_min} mid panel). In Fig.  \ref{max_min} bottom panel we show the residuals after the subtraction of the best linear fit to the ratio of the three maximum and minimum spectra. }

\subsection{{Rotational Period Estimates}}

\begin{figure*}
\centering
  \includegraphics[width=\linewidth]{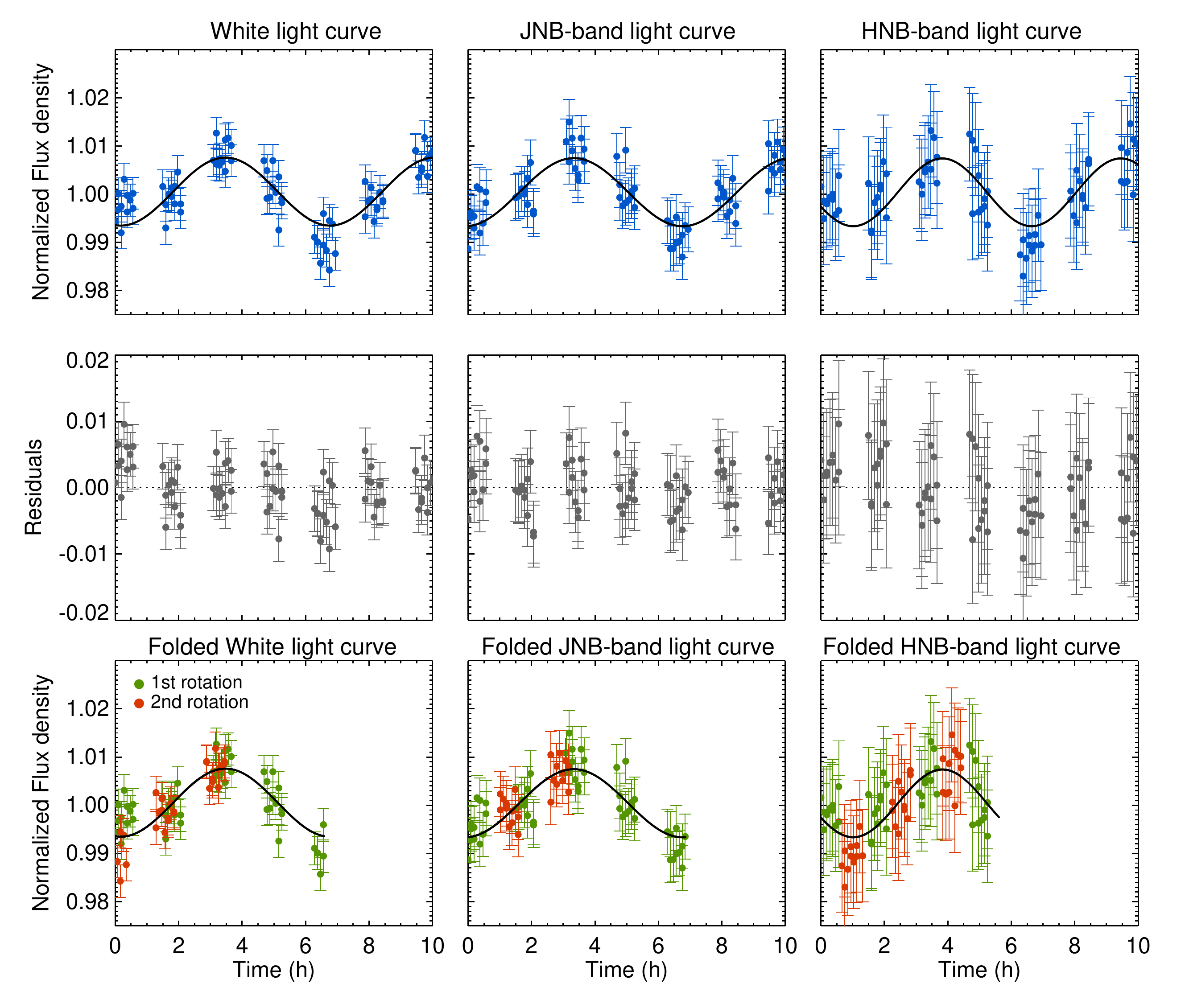}
  \caption{The upper panels show the best-fit sine function to the white, narrow $J$-band, and the narrow $H$-band light curves. Panels in the middle row show the residuals after subtracting the best-fit sine wave from the white, $J$- and narrow $H$-band light curves. The bottom panels show the respective phase-folded light curves.  }
  \label{sinefit_LC}
\end{figure*}

We aim to estimate the rotational period of Ross~458C by using its white light curve acquired in seven consecutive HST orbits, {with an end-to-end} duration of approximately 10~h.  We use three different methods to estimate the tentative rotational period of Ross~458C. {We assume that the quasi-periodic modulations we detected in its light curve are a good initial measure of the rotation period (see \citealt[][]{Apai2017}). In addition, we assume that the variability is due to rotational modulations of a hemisphere-integrated signal emerging from an atmosphere with heterogeneous cloud cover, which is the most plausible possibility and consistent with the evidence presented in the literature (see, for example, the discussion in Section 5 of \citealt{Manjavacas2018})}.

We estimate the rotational period of ROSS~458C using the three following methods:

\begin{enumerate}

\item Sine function fitting: we fit a sine function to the {all three light curves}. We chose a sine wave as it is the simplest description of the light curve, but we note that,  the actual light curve might be more complex.
We fit a sine function using the following expression {to each of these three spectral regions}:

\begin{equation}\label{sinefunction}
F(t) = A_{0} + A_{1} \sin\Big[\frac{2\pi t}{P} + \phi\Big]
\end{equation}

where $A_{0}$ is the base level of the light curve, $A_{1}$ is the amplitude $t$ is the time, $P$ is the period of the sine wave, and $\phi$ its phase.

We performed a Levenberg-Marquardt least-squares fit to the sine function using the  \texttt{mpfitfun.pro} IDL function \citep{Markwardt2008}. {This IDL function provides an estimate of the uncertainties using the uncertainties of the light curve data points, and by minimizing the value of the $\chi^{2}$}. We obtained {consistent periods for the white light curve (6.59$\pm$0.21~h), and the narrow $J$-band light curve (6.86$\pm$0.30 h), but not for the narrow $H$-band light curve (5.62$\pm$0.38~h)}. We  obtained {consistent} phases for the white (1.40$\pm$0.18~rad) and narrow $J$-band light curve (1.65$\pm$0.23~rad), but {a marginally significant different phase (2.5$\sigma$)} for the narrow $H$-band light curve (0.44$\pm$0.42~rad). Thus, there is an indication of a phase shift between the $J$- and the $H$- narrow bands, at a 2.4$\sigma$ confidence level. The residuals after the subtraction of the best-fit sine wave follow a Gaussian distribution for the white and $J$-band light curves, but not for the narrow $H$-band light curve. This indicates that there might be underlying (non-sinusoidal) structures in the narrow $H$-band light curve. {In fact, under the assumption of the light curve being a sine wave with the same period in both bands, there is no evidence for a phase shift, with phases of 1.66$\pm$0.12~rad for the $JNB$ light curve and 1.73$\pm$0.26~rad for $HNB$ light curve.}

Table~\ref{sine_fit_ampl} lists the parameters of the best-fit sine functions for the white, and the $J$- and narrow $H$-band light curves. In the top panels of Figure~\ref{sinefit_LC} we show the best-fit sine functions for the white, narrow $J$- and $H$- band light curves. In the middle row panels we show the light curve residuals after the subtraction of the best-fit sine functions. The bottom panels show the white, and narrow $J$- (JNB) and $H$-band (HNB) phase-folded light curves.

\item Bayesian Generalized Lomb-Scargle Periodogram (BGLS): As discussed in \cite{Manjavacas2018}, the regular Lomb-Scargle periodogram \citep{Horne_Baliunas1986} does not consider data point uncertainties. It has been shown that this periodogram analysis provides biased results in case of temporal gaps in the data, such as in the case of HST data \citep{Mortier2015,Cowan2017}. To solve this problem, \cite{Mortier2015} presented a Python-based code to calculate the BGLS periodogram of time-series based on algorithms presented in \cite{Bretthorst2001} and \cite{Zechmeister2009}. These consider uncertainties of the data points, data gaps, allow for zero point differences in data collected at different epochs, and also provide the probabilities of peaks of similar power in the conventional Lomb-Scargle Periodogram. With this approach we obtained a period of 6.60$\pm$0.89~h for the white light curve, a period of 6.86$\pm$0.86~h for the narrow $J$-band light curve, and a period of 5.63$\pm$1.37~h for the narrow $H$-band light curve. {Uncertainties are computed as the FWHM of a Gaussian function fitted to the peaks.}

\item Monte Carlo simulation: as a third approach we used a Monte Carlo {method} as in \cite{Manjavacas2018} to robustly estimate the periods and their uncertainties. We created 1,000 synthetic light curves based on the {observed white, JNB and HNB light curves}, and shifted each data point using normally distributed random values, using the mean of the Gaussian as the measured flux, and its standard deviation as its uncertainty. We  produced the traditional Lomb-Scargle periodogram of each simulated light curve. We calculated the period as the peak of the 50\% percentile curve, and its uncertainty as the Full Width High Maximum (FWHM) of that percentile {as in \cite{Manjavacas2018}}. We obtained a period of 6.75$\pm$1.58~h for the white light curve, a period of 6.91$\pm$1.51~h for the narrow $J$-band light curve, and a 6.27$\pm$1.55~h period for the narrow $H$-band light curve, in agreement with the periods obtained using the previous methods. In this case, we obtained the regular Lomb-Scargle periodogram because the power of the peaks obtained are normalized to the variance of the data,  {which is similar} for each generated light curve. 

\end{enumerate}

{As explained above, the lightcurve might not be, in fact, a sine wave but a more complex function (consistent with high quality lightcurves obtained for other objects). However, our current data do not warrant more complex models. Thus, we decided to adopt as a likely period, the one obtained in the  Monte Carlo simulation, 6.75$\pm$1.58~h, with its more conservative uncertainty.}

\begin{table*}
	\caption{Parameters for the sine function fit.}  
	\label{sine_fit_ampl}
	\centering
	\begin{center}
		\begin{tabular}{llll}
        \hline
			\hline 
            
 Sine component & white fit  & $JNB$-band fit & $HNB$-band fit \\   
 
 \hline
  Period (h) 	& 6.59$\pm$0.21	 & 6.86$\pm$0.30	& 5.62$\pm$0.38 \\
  $A_{0}$       & 1.00$\pm$0.01   & 1.00$\pm$0.01	& 1.00$\pm$0.01\\
  $A_{1}$	    & [-7.06$\pm$0.52]x$10^{-3}$ & [-7.06$\pm$0.74]x$10^{-3}$ &	[-7.03$\pm$1.56]x$10^{-3}$\\
  Phase	(rad)        & 1.40$\pm$0.18	 & 1.65$\pm$0.23  &	0.44$\pm$0.42\\
  $\chi^{2}$        &       93.1            &       45.0       &      17.9          \\   
  Number data points       &       77            &      77         &     77          \\   
			\hline			
		\end{tabular}
	\end{center}

\end{table*}

\section{Discussion}

\subsection{Sulfide clouds in Ross~458C}

{\cite{Burgasser2010b}, and \cite{Burningham2011} compared the near-infrared Ross~458C spectra taken with the {Folded-port Infrared Echelle (FIRE)} instrument at Mallegan, and the spectra obtained with the Infrared Camera and Spectrograph (IRCS) at Subaru,  and its mid-infrared IRAC photometry to several cloudless and cloudy atmospheric models  from \cite{Saumon_Marley2008} and from \cite{Allard2001} (BT-Settl atmospheric models).} {Both studies}  concluded that the cloudy BT-Settl models provide a significantly better match to the data than the cloud-free atmosphere models.

\cite{Morley2012} generated new atmospheric models based on the \cite{Ackerman_Marley2001} cloud models for brown dwarfs and planetary-mass objects of temperature between 400 and 1,200 K, log~$g$ = 4.0 to 5.5, and condensate efficiencies, $f_{eff}$ from 2 to 5. In agreement with \cite{Burgasser2010b}, and \cite{Burningham2011}, they concluded that cloudy atmospheric models matched significantly better the near-infrared spectrum of Ross~458C than cloud-free models. Clouds would also explain the red color of Ross~458C. \cite{Morley2012} proposed that the emergence of the sulfide clouds, specifically $\mathrm{Na_{2}S}$, might be a more natural explanation for the cloudy atmospheres predicted for Ross~458C, than the re-emergence of silicate clouds, as \cite{Burgasser2010b}, and \cite{Burningham2011} proposed.

{The discovery of rotational modulations in the near-infrared HST/WFC3 spectra in our seven HST orbits also points to the existence of clouds in the late-T Ross~458C. { This finding is consistent with the conclusions of  \cite{Morley2012}, that predicted the presence clouds of sulfide composition}}. 
Our observations of near-infrared modulations in Ross~458C extends temporally resolved spectral studies of clouds toward the very end of the {T spectral type sequence}. The detected modulations are {inconclusive} in the $HNB$, but displays a strongly wavelength-dependent slope in the $J$-band that has not yet been described in the literature. If confirmed, this color-dependent amplitude may suggest  differences in grain size distributions between silicate clouds (probed in early T-dwarfs) and sulfide clouds probably observed in the late T and Y-dwarfs. The presence of the rotational modulations in a T8 dwarf itself  suggests that clouds are typical to {most} ultracool atmospheres, although the composition and structure of the clouds is likely to differ.

\subsection{Comparison with 2MASS 22282889--4310262}

\begin{figure}
  \includegraphics[width=\linewidth]{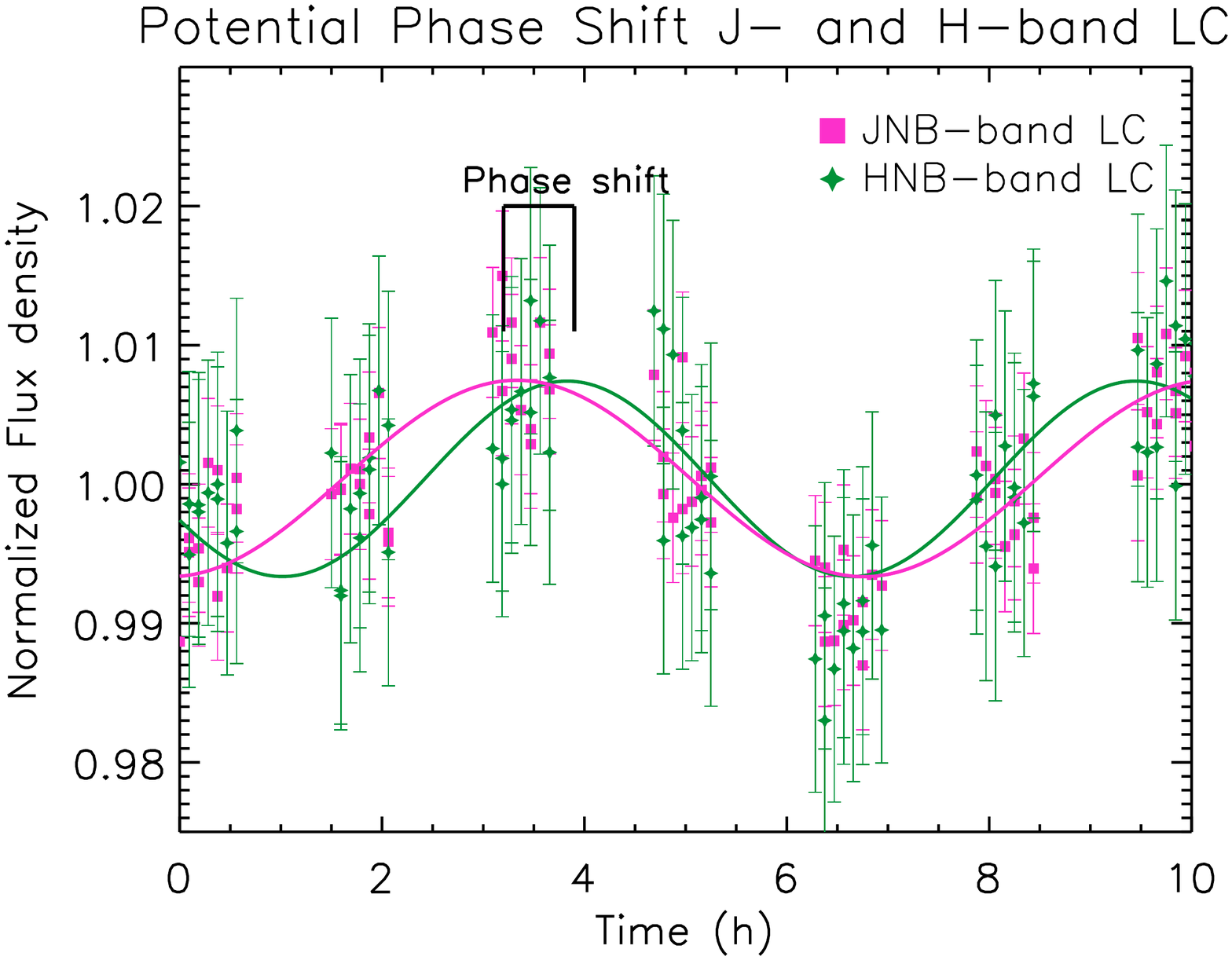}
  \caption{Narrow $J$- (pink points) and $H$-band (green points) light curves with their respective best fitting sine waves color coded with the same colors as their respective light curves.}
  \label{phase_shift}
\end{figure}

Both 2MASS 22282889--4310262 and Ross~458C show quasi-sinusoidal light curves, but with different rotational periods of $\sim$1.42~h \citep{Buenzli2012} and $\sim$6.75 h, respectively. Both $JNB$ and  $HNB$ light curves show phases shift of 1.65$\pm$0.23 rad and 0.44$\pm$0.42~rad, that correspond to 1.80$\pm$0.25 h, and 0.39$\pm$0.38 h. The likely phase shift is 1.41$\pm$0.45~h, as shown in Figure \ref{phase_shift}, {when measured at the first lightcurve's peak)}. {In contrast, for 2MASS 22282889--4310262 \cite{Buenzli2012} found that the phase shifts correlated with the pressure, which was interpreted as evidence for large-scale, longitudinal-vertical structures. Several years later follow-up observations by \cite{Yang2016} demonstrates that the phase shifts were still present, arguing for either long-lived or frequently occurring longitudinal-vertical cloud structures. If similar phase shift is confirmed in Ross 458C, it would suggest that such vertical-longitudinal structures are not rare in late-T-type brown dwarfs. \cite{Buenzli2012} measured a phase shift of 15$\pm$2 deg (0.26$\pm$0.03 rad) for 2MASS 22282889--4310262 between the $JNB$ and the $HNB$. In addition, in their two visits \cite{Yang2016} has measured  phase shifts  of  $−5\pm2$~deg ($-0.08\pm0.03$ rad), and $−8\pm2$~deg ($-0.14\pm0.03$ rad). Finally, Ross~458C has a rotational modulation amplitude of 2.62$\pm$0.02\%, and 2MASS 22282889--4310262 has a rotational modulation amplitude between 1.45\% and 5.30\% as measured by \cite{Buenzli2012}}.

\cite{Buenzli2012} fitted several atmospheric models to the spectrum of 2M2228, finding a best match with the  $\mathrm{T_{eff}}$ = 900~K, log~g = 4.5, $f_{\rm sed}$ = 3 for the \cite{Marley2012} models with sulfide clouds. Similarity, \cite{Morley2012} found a best match with sulfide clouds atmospheric models of $\mathrm{T_{eff}}$ = 700~K, log~g = 4.0, $f_{\rm sed}$ = 3 for Ross~458C. These two examples support the prediction by \cite{Morley2012} that sulfide clouds layers with patchy clouds that  condensate at temperatures cooler than 900~K could be found in late-T brown dwarfs/planetary-mass objects, and be the cause of their spectrophotometric variability.

\section{Conclusions}

\begin{enumerate}

	\item We discovered rotational modulations in the spectrum of the planetary-mass object Ross~458C. This source is the latest spectral type object for which spectral variability has been found.
    
    \item We detected rotational modulations with a {peak-to-peak} signal of 2.62$\pm$0.02\% over the entire 1.1-1.64~$\mu$m wavelength range.

	\item {Considering the results given by the method that provides the most conservative uncertainties}, we find very similar rotational periods for the white (6.75$\pm$1.58 h) and $JNB$ light curves (6.91$\pm$1.51 h), and for the $HNB$ light curve (6.27$\pm$1.55 h). 
    
    \item {Allowing different periods in the $JNB$ and $HNB$ light curves, we found a potential (2.5$\sigma$ level) phase shift between the $JNB$ (1.65$\pm$0.23 rad) and the $HNB$ light curves (0.44$\pm$0.42 rad), corresponding to 1.41$\pm$0.45~h {when measured at the first lightcurve's peak)}. A phase shift has also been detected in the T6.5 brown dwarf 2M2228 \citep[][]{Buenzli2012,Yang2015}. However, under the assumption that the lightcurve is a sine wave with the same period in both bands, there is no evidence for a phase shift. Follow-up observations are required to confirm this tentative phase shift in Ross 458C.}
    
    \item The ratio of the three maximum and the three minimum spectra across the 1.1-1.64~$\mu$m wavelength range shows a  {color trend} -- not seen previously -- in the narrow $J$-band. However, the modulations are gray in the $H$-narrow band. 
    
\end{enumerate}

 The detection of rotational modulations in Ross~458C extends detailed cloud studies to the coolest T dwarfs (T8), offering an opportunity for follow-up observations to study clouds potentially composed of sulfides. Furthermore, Ross 458C further increases the  small group of {directly imaged} planetary-mass companions, where cloud properties and rotational modulations can be studied.

\acknowledgments
Based on observations made with the NASA/ESA Hubble
Space Telescope, obtained at the Space Telescope Institute, which is operated by AURA, Inc., under NASA contract NAS 5-26555, under  GO-14241. This publication makes use of data
products from the Two Micron All Sky Survey, which is a joint
project of the University of Massachusetts and the Infrared
Processing and Analysis Center/California Institute of Technology,
funded by the National Aeronautics and Space
Administration and the National Science Foundation.


\bibliographystyle{apj}
\bibliography{ROSS458C_variability}
\end{document}